\documentclass[useAMS,usenatbib]{mnras}
\usepackage{natbib}
\usepackage{graphicx}
\usepackage{times}
\usepackage{amssymb, amsmath}
\usepackage{epstopdf}
\usepackage{color}
\usepackage[section]{placeins}
\usepackage[english]{babel}
\usepackage{lineno}
\usepackage{url}
\usepackage{ulem}
\usepackage{hyperref}
\newcommand{\kms}{\,{\rm km}\;{\rm s}^{-1}}
\newcommand{\hinvMpc}{h^{-1} {\rm Mpc}}
\newcommand{\hinvMsun}{h^{-1} {\rm M_\odot}}
\newcommand{\vecr}{\mathbf r}

\def\aap{A\&A}

\def\apj{ApJ}
\def\apjs{ApJS}
\def\apjl{ApJL}

\def\mnras{MNRAS}
\def\aj{AJ}

\title[Clustering of Faint Red Galaxies]{On the Clustering of Faint Red Galaxies}
\author[Xu et al.]{
Haojie Xu$^1$\thanks{E-mail: haojie.xu@utah.edu}, 
Zheng Zheng$^1$, 
Hong Guo$^{2,1}$,
Ju Zhu$^3$,
and
Idit Zehavi$^4$\\
$^1$
Department of Physics and Astronomy, University of Utah, 115 South 1400 East,
Salt Lake City, UT 84112, USA\\
$^2$
Shanghai Astronomical
Observatory, Chinese Academy of Sciences, Shanghai 200030, China\\
$^3$
Department of Astronomy, University of Illinois at Urbana-Champaign,
MC-221, 1002 W. Green Street, Urbana, IL 61801, USA\\
$^4$
Department of Astronomy, Case Western Reserve University, 10900 Euclid Avenue,
Cleveland, OH 44106, USA\\
}
\begin{document}
\maketitle

\begin{abstract}
Faint red galaxies in the Sloan Digital Sky Survey show a puzzling clustering 
pattern in previous measurements. In the two-point correlation function (2PCF),
they appear to be strongly clustered on small-scales, indicating a tendency to
reside in massive haloes as satellite galaxies. However, their weak 
clustering on large scales suggests that they are more likely to be found in 
low mass haloes. The interpretation of the clustering pattern suffers from the
large sample variance in the 2PCF measurements, given the small volume of 
the volume-limited sample of such faint galaxies. We present improved 
clustering measurements of faint galaxies by making a full use 
of a flux-limited sample to obtain volume-limited measurements with an 
increased effective volume. In the improved 2PCF measurements, the fractional
uncertainties on large-scales drop by more than 40 per cent, and the strong contrast
between small-scale and large-scale clustering amplitudes seen in previous 
work is no longer prominent. From halo occupation distribution modelling of the
measurements, we find that a considerable fraction of faint red galaxies 
to be satellites in massive haloes, a senario supported by the strong
covariance of small-scale 2PCF measurements and the relative spatial 
distribution of faint red galaxies and luminous galaxies. However, the 
satellite fraction is found to be degenerate with the slope of the 
distribution profile of satellites in inner haloes. We compare the
modelling results with semi-analytic model predictions and discuss the 
implications.
\end{abstract}


\section{Introduction}
\label{sec:intro}

Galaxy clustering measured in large galaxy redshift surveys encodes important
information about galaxy formation and evolution. In particular, galaxy 
clustering measurements, in combination with the theoretically well-understood
dark matter halo population for a given cosmology, enable the inference 
of the relation between galaxies and dark matter haloes, with halo-based 
clustering model \citep[e.g.][]{Jing98,Peacock00,Seljak00,Scoccimarro01,
Berlind02,Yang03,Zheng05}. In this work, we focus on measuring and modeling
the clustering of a population of faint red galaxies and discuss the 
corresponding implications on their formation and evolution.

Galaxies occupying the faint, red corner of the galaxy colour-magnitude 
diagram display interesting properties. Using Sloan Digital Sky Survey 
(SDSS; \citealt{York2000}) data, \citet{Hogg03} and \cite{Blanton05a} find
that both the faint and bright red galaxies have the highest galaxy 
over-densities within spheres of radii $1\hinvMpc$ and $8\hinvMpc$, indicating
the high density environments they reside in. \cite{Norberg02} analyse
the 2dF Galaxy Redshift Survey data and concludes that the faint, early-type
galaxies has nearly the same clustering strength with bright ones, which
contrasts the idea that brighter galaxies usually have stronger clustering
strength. With the early SDSS data, \citeauthor{Zehavi05} (\citeyear{Zehavi05}; 
hereafter Z05) find that the two-point correlation function (2PCF)
of faint red galaxies rises steeply toward small
scales ($\la 2\hinvMpc$). In particular, the feature is prominent in the
faintest sample ($-19<M_r<-18$) they analysed. In addition, the 2PCF of 
this sample shows 
a quite low clustering amplitude on large scales ($\ga 2\hinvMpc$). These
features are not seen in the blue galaxies. The
result is further confirmed with the analysis of the data from the completed
SDSS survey (\citealt{Zehavi11}; hereafter Z11) --- the small-scale clustering
amplitude is as high as that of the brightest sample ($\sim$4 magnitudes more
luminous), in strong contrast to the low large-scale clustering amplitude. The
case is even more extreme for a fainter sample with $-18<M_r<-17$ (whose
2PCF measurements have a low signal-to-noise ratio though). 

The strong non-linearity in the 2PCF towards small scales, together with the
small-scale redshift-space distortion, can explain the scale- and 
method-dependent galaxy bias factor determined for the faint red galaxies 
from many previous studies 
\citep[e.g.][]{Li06a,Swanson08,Cresswell09,Zehavi11}, 
as investigated in detail by \citet{Ross11}.

The strong small-scale clustering indicates that the faint red galaxies 
tend to reside in massive haloes, while the weak large-scale clustering 
suggests that they are more likely to be found in 
low mass haloes. This puzzling result leads to some difficulty in modeling 
the 2PCF within the HOD framework. The simple HOD model in Z11 results in a 
satellite fraction of 90 per cent, with a poor fit. The other model Z11 considered 
is that the $-19<M_r<-18$ faint red galaxies are composed of central galaxies in
haloes of a few times $10^{11}\hinvMsun$ and satellite galaxies in haloes
more massive than $10^{13}\hinvMsun$, with no galaxies in host haloes of mass 
in between. The model leads to a satellite fraction of 34 per cent, but it still 
cannot provide a good explanation of the low amplitude at large scales. 
Other statistics are also used to study faint red galaxies. The 
red galaxy pairwise velocity dispersion on small scales peak around 
$M_r=-19.5$ \citep{Li06b}, which supports the scenario that a fraction of 
the faint red galaxies are satellites in massive haloes. With a group catalog, 
\citet{Wang09} study red galaxies with much lower luminosity and infer that 
about 55--65 per cent of $M_r\sim -17$ red galaxies are satellites.

Different processes may be at work to quench star formation in central and
satellite galaxies, causing them to turn red. If the faint red galaxies are
central galaxies in low mass haloes ($\sim 10^{11}\hinvMsun$), star formation
feedback can be responsible for shutting down the star formation. If the faint 
red galaxies are satellites in massive haloes, environmental effects (e.g. 
strangulation, ram pressure stripping, and tidal heating and stripping) are
the more likely causes. A better determination of the central and satellite
status of the faint red galaxies can help to constrain their formation.
Using galaxy clustering for such a study relies on accurate clustering
measurements. However, given the low luminosity of the faint red galaxies, 
they cannot be observed to large distances in a flux-limited galaxy redshift 
survey, like the SDSS. The clustering measurements, which are usually done 
with volume-limited samples, 
suffer from substantial sample variance with the small survey volume for faint
galaxies. One way to overcome the limitation is to study the angular 
clustering with the (deeper) photometric sample \citep[e.g.][]{Ross11}. In 
this paper, we focus on a maximal use of the spectroscopic sample to improve
the clustering measurements of faint galaxies and investigate what constraints
we can put on their nature through HOD modeling.

In Section~\ref{sec:data}, we describe the data and clustering measurements.
In Section~\ref{sec:pairwise}, we present the method to measure the 2PCF 
in a volume-limited sense with the full use of a flux-limited sample, 
improving the measurements by effectively increasing the volume. 
A detailed derivation of the method is given in Appendix~\ref{sec:appendix1}. 
The results from HOD modeling of the improved 2PCF is presented in 
Section~\ref{sec:hod}. Finally, we summarize and discuss the results in 
Section~\ref{sec:discussion}.

\section{Observations and Methods}
\label{sec:data}

\subsection{Data}

We study the faint red galaxies in the MAIN sample \citep{strauss02} of the
SDSS \citep{York2000,stoughton02}. The data we 
use are from the SDSS Data Release 7 (hereafter SDSS DR7; 
\citealt{Abazajian09}), which is the end data release of SDSS-II.  
We take the large-scale structure sample in the NYU Value-Added Galaxy 
Catalog\footnote{\url{http://sdss.physics.nyu.edu/lss.html}} 
(NYU-VAGC; \citealt{Blanton05b,Adelman08,Padmanabhan08}),
which covers a sky area of 7,966 deg$^2$.
We adopt the full flux-limited sample, and in Section~\ref{sec:pairwise}, 
we show how to use the sample in a pairwise volume-limited sense.
The magnitude in the sample has been $K$-corrected and we further apply an 
evolution correction to $z\sim 0.1$ (Z11). Galaxies with 
$-19<M_r<-18$ are selected from the NYU-VAGC bright sample with $r<17.6$. 
Finally we define our faint red galaxy sample by imposing the colour
selection $ g-r > 0.21-0.03M_r$, as in Z11. In total, we end up with 
11,357 faint red galaxies. 

For the clustering measurements, we also use the  corresponding random catalogs
provided by the NYU-VAGC, which takes accounted for the complex survey 
geometry and detailed angular selection function. Compared to the galaxy
catalog, the random catalog we use contains 50 times as many galaxies.

Throughout the paper, all distances and galaxy pair separations are expressed
in comoving units. For all distance calculations, we assume a flat $\Lambda$CDM
cosmological model with $\Omega_m = 0.25$. Since the sample is at $z\sim 0$, 
the results are not sensitive to this choice of $\Omega_m$. The 
Hubble constant is taken to be $H_0=100h \kms {\rm Mpc}^{-1}$ with $h=0.7$, 
while the absolute magnitudes of galaxies are expressed by adopting $h=1$.

\subsection{Methods}

We characterize the clustering of galaxies with the 2PCF, the excess 
probability of finding galaxy pairs at a given separation with respect to 
a random distribution \citep{Peebles80}. For a pair of galaxies with 
redshift positions $\mathbf v_{1}$ and $\mathbf v_{2}$, we compute the 
separation ($\pi$) parallel to the line of sight and that ($r_p$) perpendicular
to the line of sight as
\begin{equation}
\pi \equiv |\mathbf s \cdot \mathbf l|/|\mathbf l|,\,\,  r_p^2 \equiv \mathbf s \cdot \mathbf s - \pi^2,
\end{equation}
where $\mathbf s \equiv \mathbf v_1 - \mathbf v_2$ is the redshift 
separation vector and $\mathbf l \equiv (\mathbf v_1 + \mathbf 
v_2)/2$ is the line-of-sight vector \citep{Fisher94}.

The widely adopted estimator to calculate the redshift-space galaxy 2PCF 
$\xi(r_{p},\pi)$ is given by \cite{LandySzalay93},
\begin{equation}
\label{eqn:LS}
\xi(r_p,\pi)=\frac{{\rm DD}-2{\rm DR}+{\rm RR}}{\rm RR},
\end{equation}
where DD, RR, and RR are the numbers of data-data, data-random and
random-random pairs in each separation bin, normalized by the corresponding
total pair counts $N_{\rm D}(N_{\rm D}-1)/2$, $N_{\rm D}N_{\rm R}$ and 
$N_{\rm R}(N_{\rm R}-1)/2$ (with $N_{\rm D}$ and $N_{\rm R}$ the total numbers
of galaxies and random points), respectively. In Section~\ref{sec:pairwise}, 
we propose to measure the 2PCF using the flux-limited sample but with 
appropriate volume weight for each galaxy pair to reach an effectively 
volume-limited measurement. The method can make full use of the observation 
and measure the 2PCF more accurately. The generalized estimator is 
presented in Appendix~\ref{sec:appendix1} and discussed in 
Section~\ref{sec:pairwise}.

From the measured $\xi(r_p,\pi)$, we compute the projected 2PCF
\begin{equation}
w_p(r_p)=2\int_0^\infty d\pi\xi(r_p,\pi),
\end{equation}
which intends to remove the redshift-space distortion effect. For the 
faint red galaxy
sample studied in this paper, we integrate up to $\pi_{\rm max} = 40\hinvMpc$,
which is also adopted consistently in the HOD modeling of the clustering 
result. In our measurements, we use linearly space bins in $\pi$ with widths 
of $1\hinvMpc$ and logarithmically spaced $r_p$ bins with 0.2 dex width.

Finally, to estimate the covariance matrix of the measured $w_p$, we apply
the jackknife resampling method (e.g., Z05, Z11) with $N=146$ spatially 
contiguous subsamples of the full data set with equal area. We measure the 
projected 2PCF $w_p^l$ for the $l$-th jackknife sample ($l=1,2,...,N$), which 
omits the $l$-th subsample, and the covariance matrix is then estimated as
\begin{equation}
{\rm Covar}(w_{p,i},w_{p,j}) = \frac{N-1}{N}\sum\limits_{l=1}^N 
(w_{p,i}^l-\bar{w}_{p,i})(w_{p,j}^l-\bar{w}_{p,j}),
\end{equation}
where $w^l_{p,i}$ is the value of $w_p$ at the $i$-th projected pair 
separation bin $r_{p,i}$ measured from the $l$-th jackknife sample, and 
$\bar{w}_{p,i}$ is the value of $w_p$ in such a bin averaged over all 
jackknife samples.
 
\section{Pairwise Volume-limited 2PCF measurements from the Flux-limited Sample}
\label{sec:pairwise}

\begin{figure*}
\includegraphics[width=1.0\textwidth]{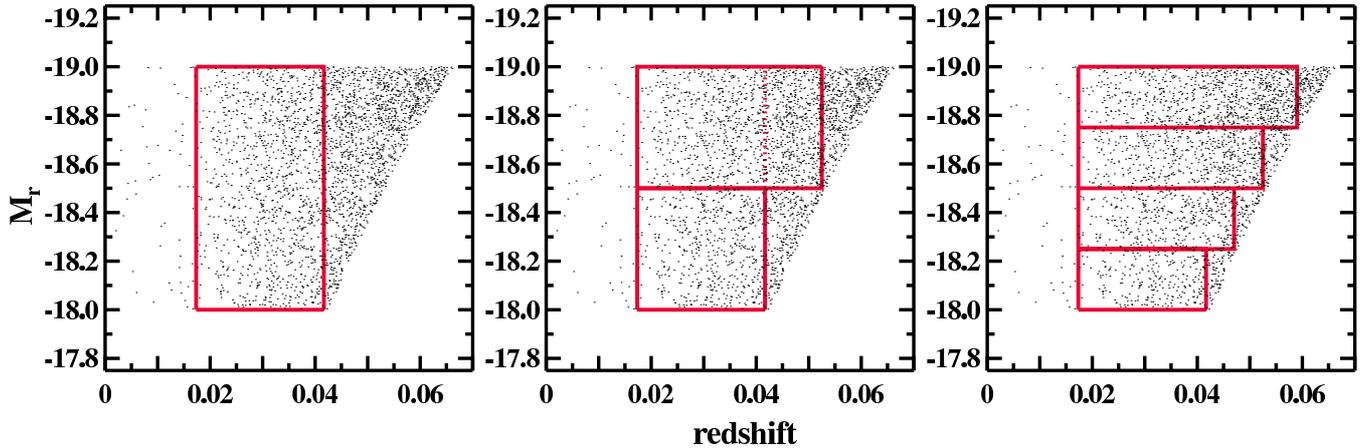}
\caption{
Illustration of different ways of sample construction in measuring the 2PCF 
of red galaxies in the luminosity bin of $-19<M_r<-18$ in a volume-limited 
sense. 
In each panels, the minimum redshift $z_{\rm min}=0.017$, delineating by the
left vertical line. The dots show a random selection of the flux-limited 
sample, with the right boundary set by the flux limit.
{\it Left:} The commonly adopted volume-limited sample, with the maximum 
redshift (the right side of the red rectangle) determined by the faint boundary
of the luminosity bin.
{\it Middle:} Galaxies are divided into two
subsamples.
The contribution to the 2PCF from the bright galaxies can be measured 
within the larger volume (top rectangle), reducing the effect of sample 
variance. 
{\it Right:} Galaxies are divided into four
subsamples, with different volumes. More galaxies and an effectively larger 
volume are used in measuring the 2PCF. 
The method introduced in this work
can be thought as dividing galaxies into subsamples of 
infinitesimally small luminosity bins, achieving a maximal use of the 
flux-limited data for an effectively volume-limited 2PCF measurement.
See the text for details.
}
\label{fig:illustration}
\end{figure*}

In large galaxy redshift surveys, volume-limited galaxy samples are preferred 
for galaxy clustering study (e.g., Z05, Z11), since the samples are 
well-defined within the volume to have uniform properties (e.g., the
same range of luminosities across the volume). The measured 2PCFs from 
volume-limited samples also make it easy to interpret with theoretical models. 
For example, with HOD modeling of the clustering of volume-limited samples, 
we can infer meaningful information about the relation between galaxy 
properties (e.g., luminosity) and halo mass. 

However, the natural product from a galaxy redshift survey is a flux-limited 
galaxy sample, which is substantially larger than any volume-limited sample 
constructed for galaxy clustering study. The volume (or maximum redshift) for 
a volume-limited sample is determined by requiring that all type of galaxies 
in the sample to be complete within the volume. For some galaxies, even though
their properties can make them fall into the sample of interest, they are 
excluded because they fall outside of the volume of completeness. 
This greatly limits the 
statistical power of the volume-limited sample. To have better statistics for 
the clustering of galaxies, it is desirable to include as many galaxies as 
possible while still keeping the volume-limited nature of the 2PCF 
measurements. This is particularly important for faint galaxies, given
the expected small volume for constructing a full volume-limited sample.
We propose here a method of measuring galaxy 2PCFs by making full use of
galaxies in a flux-limited sample and giving each pair of galaxies an
appropriate weight, which will lead to effectively volume-limited 
2PCF measurements. The method is termed as pairwise volume-limited
2PCF measurements, whose meaning will become clear later.

We use the faint red galaxy sample to illustrate the basic idea of the 
method and present the final pairwise volume-limited 2PCF measurements to be 
used for the HOD modeling.

In the left panel of Figure~\ref{fig:illustration}, we show the construction
of the volume-limited sample for $-19<M_r<-18$ red galaxies, as used in
previous studies (e.g. Z11). The points denotes a random selection of
the red galaxies within
the $-19<M_r<-18$ magnitude bin that are observed in the SDSS-II DR7, which
are flux limited. The right boundary of the points reflects the flux limit or 
the faint magnitude limit $r = 17.6$. To construct the volume-limited sample,
both the galaxies at the faint end ($M_r=-18$) and at the bright end 
($M_r=-19$) need to be complete, and the maximum redshift is then completely
determined by the faintest galaxies in the sample of interest (i.e. 
$M_r=-18$). As a consequence, only galaxies within the red box enter the
volume-limited sample. The maximum redshift is $z_{\rm max}=0.042$. In
principle, the minimum redshift can go to zero, but here we follow Z11
to fix it at $z_{\rm min}=0.017$ to avoid low z incompleteness, which 
only reduces the volume by $\sim$7 per cent.
The volume-limited sample constructed this way is composed of 5,165 galaxies.
It clearly excludes all the galaxies beyond $z_{\rm max}$, because they 
are not complete in the full range of the above magnitude bin.
The number of such galaxies is not small, $\sim$6,000, comparable to 
that in the volume-limited sample. If they can be used for measuring the 
2PCFs, we would achieve a better statistic.

The middle panel of Figure~\ref{fig:illustration} shows a case on how to
make use of the galaxies beyond $z_{\rm max}$, which can
be treated as an instructive example and an intermediate step towards 
the method 
we propose. We divide the $-18<M_r<-19$ red galaxies into two subsamples
--- the faint and bright subsamples are made of galaxies with $-18<M_r<-18.5$ 
and
$-18.5<M_r<-19$, respectively. The principle underlying our method is that
galaxy pairs are made of faint-faint pairs, faint-bright pairs, and 
bright-bright pairs. In terms of the 2PCF $\xi$, it can be decomposed into 
contributions from the auto-correlation $\xi_{ff}$ of galaxies in the faint 
subsample, the cross-correlation $\xi_{fb}$ between galaxies in the faint
and bright subsamples, and the auto-correlation $\xi_{bb}$ of galaxies in the
bright subsamples. The total 2PCF can be expressed as \citep{Zu08} 
\begin{equation}
\label{eqn:decomposition}
\xi = \frac{\bar{n}_f^2}{\bar{n}^2}\xi_{ff}
     +\frac{2\bar{n}_b\bar{n}_f}{\bar{n}^2}\xi_{fb}
     +\frac{\bar{n}_b^2}{\bar{n}^2}\xi_{bb},
\end{equation}
where $\bar{n}_f$, $\bar{n}_b$, and $\bar{n}$ are the galaxy number densities 
in the faint subsample, the bright subsample, and the full sample, 
respectively.

The auto-correlation $\xi_{ff}$ of galaxies in the faint subsample can be
measured with the volume-limited faint subsample, whose volume is
illustrated by the bottom red rectangle box in the middle panel of 
Figure~\ref{fig:illustration}. It has the same $z_{\rm max}$ as the case 
in the left panel. For the cross-correlation $\xi_{fb}$ between the
faint and bright subsamples, the volume-limited measurements are also
made in the above volume, i.e., the same as in the left panel. However,
for the auto-correlation $\xi_{bb}$ of galaxies in the bright subsample,
we are able to perform the measurement in a much larger volume, as
illustrated by the top (solid) red rectangle in the middle panel. It reaches
up to $z_{\rm max}=0.053$, and the galaxies in the bright sample are
complete in this volume. That is, we are able to measure $\xi_{bb}$ in a
volume almost twice larger than in the previous case, which reduces the
uncertainty (sample variance) in $\xi_{bb}$ and hence improves the 
measurement in the total
$\xi$ [equation~(\ref{eqn:decomposition})]. The measurement is still a
volume-limited one --- we do not change the component contributions to the
total $\xi$, but just make the measurement more accurate for the bright-bright
auto-correlation contribution. In other words, we increase the effective 
volume for the 2PCF measurement.

\begin{figure}
\includegraphics[width=0.45\textwidth]{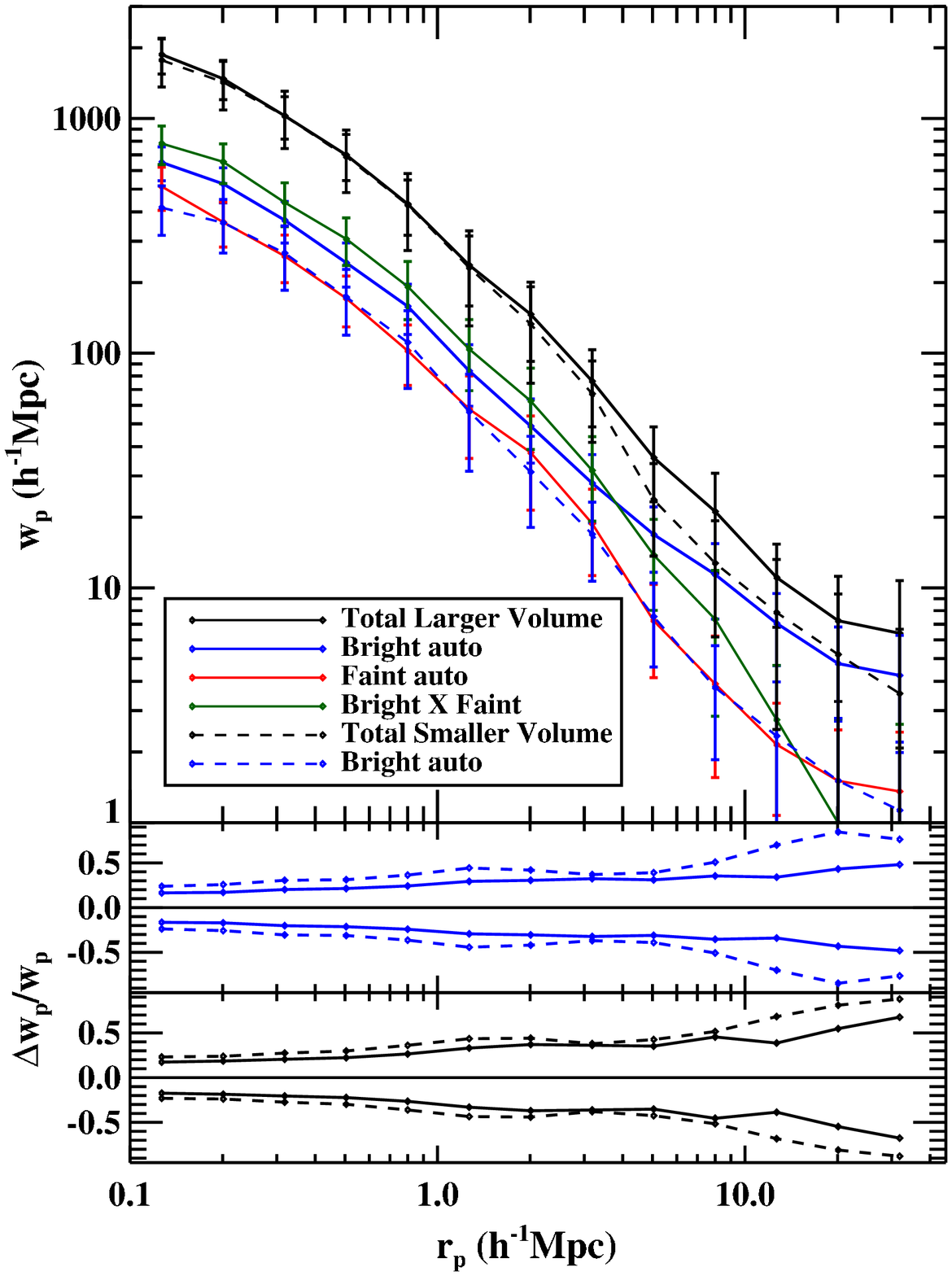}
\caption{
Decomposition of the projected 2PCF of the $-19<M_r<-18$ red galaxies with the 
sample
division illustrated in the middle panel of Figure~\ref{fig:illustration}.
With the sample divided into the faint and bright subsamples, the total
2PCF (solid black) is contributed by the faint-faint auto-correlation
(solid red), the faint-bright cross-correlation (solid green), and the
bright-bright auto-correlation (solid blue). The first two are measured in
a smaller volume with $z_{\rm max}=0.042$, while the last one in a larger
volume with $z_{\rm max}=0.053$ (Fig.\ref{fig:illustration}). For comparison,
the total and the bright-bright 2PCF measured in the smaller volume are also
shown (dashed black and blue), corresponding the the case with the commonly
adopted volume-limited sample illustrated in the left panel of 
Figure~\ref{fig:illustration}. The middle and bottom panels compare the 
fractional uncertainties in the bright-bright and total 2PCFs, respectively.
}
\label{fig:xifaintbright}
\end{figure}

Figure~\ref{fig:xifaintbright} shows the measurements of different component
2PCFs and compare the results corresponding to the case in the middle panel
with the usual volume-limited case in the left panel of 
Figure~\ref{fig:illustration}. For simplicity, below we use
$\xi$ to denote the clustering measurements of interest here, i.e., $w_p$. 
The components $\xi_{ff}$ 
(red curve) and $\xi_{fb}$ (green curve) are the same for the two cases.
The component $\xi_{bb}$ shows a clear difference. The $\xi_{bb}$ measured
in the larger volume (solid blue curve) is higher in amplitude, compared to 
that measured in the smaller volume (dashed blue curve). This is clearly a 
sample variance effect, and the one measured in the larger volume is less
affected. The fractional uncertainties (middle panel) estimated from the 
jackknife method also shows that the larger volume reduces the uncertainties
in the measurements. Because of the improvement in the $\xi_{bb}$ measurements,
the uncertainties in the total $\xi$ are reduced by 20-40 per cent, especially 
on large scales (bottom panel). By comparing the solid and dashed black 
curves, we find that the improved $\xi$ measurement no longer shows the 
steep drop at $r_p>2\hinvMpc$ seen in Z11. 

We see that dividing the sample into faint and bright subsamples enables
us to improve the 2PCF measurements. However, we still exclude many galaxies,
i.e. galaxies at redshifts higher than those set by the volume-limited
faint and bright subsamples. These are the points out of the right boundaries 
of the top and bottom solid rectangles in the middle
panel of Figure~\ref{fig:illustration}.

To incorporate more galaxies into the measurements, we can make further 
division for the faint and bright subsamples. That is, we divide the whole
sample into more subsamples with finer luminosity bins. 
Following \citet{Zu08}, equation~(\ref{eqn:decomposition}) is then 
generalized to
\begin{equation} 
\label{eqn:decompij}
\xi=\sum\limits_{i,j}\frac{\bar{n}_i\bar{n}_j}{\bar{n}^{2}}\xi_{ij},
\end{equation}
where $\bar{n}_i$ is the mean number density of the $i$-th subsample and
$\xi_{ij}$ is the 2-point cross-correlation function between galaxies in 
the $i$-th and $j$-th subsample (autocorrelation function if $i=j$). Both 
$i$ and $j$ go from 1 to $N$, the number of luminosity bins. We can construct
volume-limited sample for each subsample. Given the difference in luminosity, 
different subsamples have different values of $z_{\rm max}$ or volumes of
completeness. Except
for the faintest sample, the volumes are all larger than the one used in 
the left panel of Figure~\ref{fig:illustration}. When measuring $\xi_{ij}$,
we just need to use galaxies in the volume defined by the fainter one of 
the $i$-th and $j$-th subsamples. The right panel of 
Figure~\ref{fig:illustration} 
illustrates one case of dividing the sample into four subsamples, and we can 
see that the volume to be used increases with the luminosity of the subsample.
Compared to the middle panel, more galaxies are included in the 2PCF 
measurements, and we expect the measurements to be further improved, reducing
the effect of sample variance. 

In Appendix~\ref{sec:appendix1}, we prove that the 2PCF with such a division 
can be calculated with the following generalized Landy-Szalay estimator,
\begin{equation}
\label{eqn:estimator}
\xi=\frac{
  \sum\limits_{i,j}\frac{1}{V_{ij}}\left.{\rm DD}_{ij}\right\vert_{V_{ij}}
-2\sum\limits_{i,j}\frac{1}{V_{ij}}\left.{\rm DR}_{ij}\right\vert_{V_{ij}}
 +\sum\limits_{i,j}\frac{1}{V_{ij}}\left.{\rm RR}_{ij}\right\vert_{V_{ij}}
}{
  \sum\limits_{i,j}\frac{1}{V_{ij}}\left.{\rm RR}_{ij}\right\vert_{V_{ij}}
}.
\end{equation}
Here $V_{ij}={\rm min}\{V_i,V_j\}$, where $V_i$ and $V_j$ are the volumes 
of completeness corresponding to the $i$-th and $j$-th luminosity bin given 
the flux limit. The quantity $\left.{\rm DD}_{ij}\right\vert_{V_{ij}}$ is the 
normalized number of galaxy pairs within $V_{ij}$, with each pair composed of 
one galaxy in the $i$-th luminosity bin and one in the $j$-th luminosity bin,
and similarly for $\left.{\rm DR}_{ij}\right\vert_{V_{ij}}$ and 
$\left.{\rm RR}_{ij}\right\vert_{V_{ij}}$. Note that for our purpose here, each point 
in the random catalog is assigned with a luminosity so that the luminosity 
and redshift distributions of the galaxy sample are reproduced. The way we 
construct the random catalog is as follows. For each random point, we take 
the angular position in the original random catalog and assign a 
luminosity randomly chosen from the galaxy catalog. We then 
assign a redshift to this random point in the range of ($z_{\rm min}$, 
$z_{\rm max}$), where $z_{\rm min}=0.017$ and $z_{\rm max}$ is determined 
from the luminosity and the flux limit. 
The redshift is assigned such that galaxies are uniformly distributed in the
comoving volume set by $z_{\rm min}$ and $z_{\rm max}$ 
(i.e. $D_c^3$, the cube of the comoving distance, follows
a uniform distribution).

Conceptually, the ultimate limit of the division is that there is one galaxy 
per fine luminosity bin, which has a corresponding maximum volume determined 
from the luminosity and the flux limit. This is the limit that we apply 
equation~(\ref{eqn:estimator}) to compute the 2PCF. In detail, for each 
galaxy-galaxy, galaxy-random, or random-random pair, we check whether 
both galaxies fall into the volume $V_{ij}$ (which is determined by the 
fainter galaxy of the pair). If they do, we keep the pair and weigh it 
by $1/V_{ij}$. 
The generalized estimator in equation~(\ref{eqn:estimator}) can reduce 
to the same form as the Landy-Szalay estimator [equation~(\ref{eqn:LS})], 
with the DD, DR, and RR terms being understood as the total count of pairs 
that pass the above criterion with each weighted by $1/V_{ij}$. 
We note that \citet{Li06a}, \citet{Li09} and \citet{Li12} apply a 
similar $1/V$ weight (explicitly mentioned but without a derivation; with
$V$ the smaller $V_{\rm max}$ of the two galaxies) to galaxy pairs in 
estimating the distribution of stellar mass in the universe and the
stellar-mass-dependent galaxy clustering. See Appendix~\ref{sec:appendix1} for
a discussion of the subtle difference between our method and theirs.

Since we count each pair in a volume-limited sense, we call the measurements
from this method as pairwise volume-limited measurements. Clearly this 
method makes the maximal use of the data, trying to account for contributions 
from every possible galaxy pair that belong to the sample. The measurements
are still effectively volume-limited, since the component 2PCF is computed 
within the respective volume that ensures the completeness of the subsamples
of interest. Compared to the original, commonly used 
volume-limited sample, which we denote it as `faint-end volume-limited',
the new method measures the 2PCF contributions from brighter galaxies in the 
sample in a larger volume. This helps to reduce the sample variance effect,
and improve the overall 2PCF measurements. One may wonder whether the 
different volumes used to compute the component 2PCFs complicate the
estimation of the covariance matrix. In fact, the covariance matrix is 
estimated using the jackknife method, and for each jackknife sample we apply 
the same generalized estimator. Therefore the variations in the volumes of 
different subsamples are automatically accounted for in the covariance 
matrix estimation. Not surprisingly, the method shares some similarities 
with the $1/V_{\rm max}$ method used in estimating the luminosity function 
from a flux-limited galaxy survey 
--- the $1/V$ weight is used to estimate the number density of galaxies 
in deriving the luminosity
function,\citep[e.g.][]{Schmidt68,Huchra73}, 
while it is used to effectively estimate 
the number density of galaxy pairs in our method.

We emphasize that the pairwise volume-limited measurement is different from 
the flux-limited measurement, even though both are based on the flux-limited 
galaxy sample. The flux-limited measurement does not remove any galaxy pairs
and there is no weight given to each pair. In terms of the random sample, 
no luminosity information is needed, and we only need to ensure that the 
random points reproduce the angular and redshift distributions of the galaxy
sample. Since brighter galaxies are observed to larger redshifts, they 
contribute more pairs than in a faint-end volume-limited case. As a 
consequence, the flux-limited 2PCF measurement effectively gives larger 
weights to the component 2PCF contributions from brighter galaxies, making 
it hard to interpret 
(see, however, \citealt{Zu15}, who manage 
to model the flux-limit measurements to gain an improved statistical power).

\begin{figure}
\includegraphics[width=0.45\textwidth]{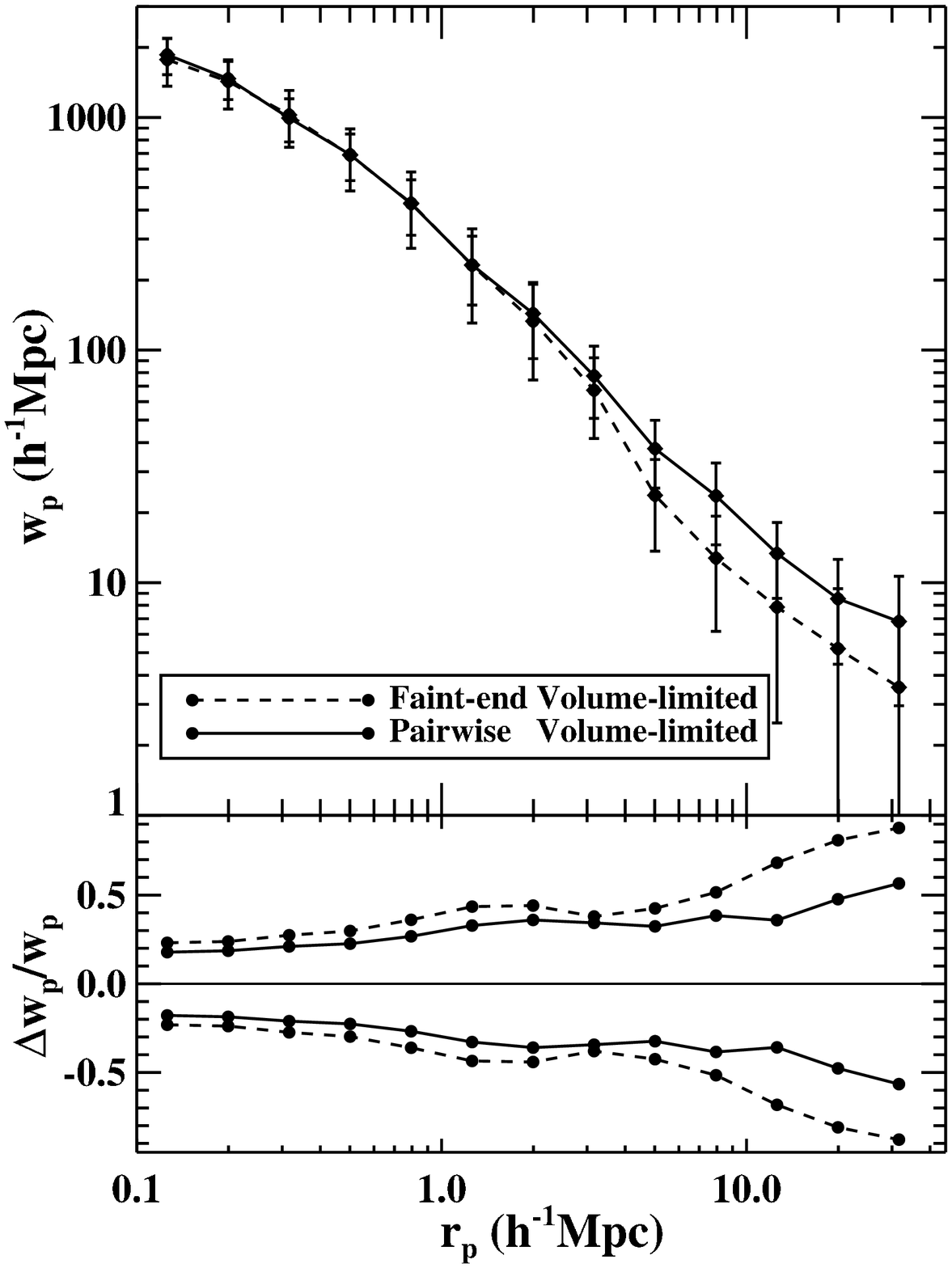}
\caption{
Projected 2PCF measurements (top) and fractional uncertainties (bottom) of 
the $-19<M_r<-18$ red galaxies from two methods. The dashed curves correspond
to the case with the commonly adopted volume-limited sample (termed as
`faint-end volume-limited' here), and the solid curves are for the case
of the pairwise volume-limited measurements with the flux-limited sample.
See text for details.
}
\label{fig:xi}
\end{figure}

Finally, we apply the pairwise volume-limited method to the faint red galaxy 
sample and
compute the 2PCF to be used in the next section for HOD modeling. In 
the top panel of Figure~\ref{fig:xi}, the sold curve show the pairwise 
volume-limited $w_p$ measurements. For comparison, the
faint-end volume-limited measurements are plotted as the dashed curve, which
are essentially the same as those obtained by Z11. Similar to the case in
Figure~\ref{fig:xifaintbright}, the full pairwise volume-limited measurements
show enhanced clustering amplitudes on large scales, by a factor of about 
60 per cent. The original finding 
by Z11 of a steep drop beyond $r_p>2\hinvMpc$ is less prominent. The bottom 
panel shows the fractional uncertainties of the measurements. Those associated
with the pairwise volume-limited measurements are smaller. Especially on large
scales, the fractional uncertainties decrease by more than 40 per cent, which 
corresponds to an effective change in survey volume by a factor of two. 
We also applied this method to an even fainter sample, the red galaxies with 
$-18<M_r<-17$, but the measurement is still noisy (like in Z11) because of 
the very limited sample size and volume.

The left panel of Figure~\ref{fig:covar} shows the covariance matrix estimated
from the jackknife samples. It shows the interesting pattern that data points 
below 3$\hinvMpc$ are highly correlated. This is in direct contrast to the
pattern seen in the covariance matrices for bright galaxy samples. As a 
comparison, the right panel shows that for the $M_r<-21$ sample, where the
positive correlation occurs for data points on scales above 3$\hinvMpc$. 
A similar behavior was also found in Z05. We verified that the pattern 
seen in the covariance matrix for the $M_r<-21$ sample remains similar, even 
if it is evaluated within a small volume. Therefore, the contrast in the 
covariance matrices for faint and 
bright galaxy samples is not caused by the difference in survey volume. It
may reflect the difference in the way that galaxies occupy dark 
matter haloes and in the contribution from satellite galaxies, which will be 
discussed further in the Section~\ref{sec:hod}.

Before switching to the HOD modeling, we point out that the pairwise 
volume-limited method works quite efficiently for faint galaxy samples with 
luminosity below $L_*$, while it leads to little improvement for luminous 
galaxy samples (above $L_*$). We have verified this by comparing the 
faint-end volume-limited and pairwise volume-limited measurements for the 
$M_r<-21$ sample. This difference between faint and bright galaxy samples 
results from the shape of the galaxy 
luminosity function. For luminosity $L>L_*$, the number density of galaxies
drops exponentially towards the high luminosity end. Since the 2PCF component 
contribution from galaxies depends on their number density 
[equation~(\ref{eqn:decompij})], the steep drop in the number density of
brighter galaxies means that they only make a smaller contribution to the
total 2PCF. Even their higher clustering amplitude cannot compensate the
rapid (exponential) drop in their number density. For $L<L*$ galaxies, the 
number density change is not as steep (power-law rather than exponential),
and the brighter galaxies in the sample can make a substantial contribution 
to the overall 2PCF. As a consequence, a better measurement of the clustering 
of brighter galaxies with the pairwise volume-limited method improves the 
overall 2PCF measurement.

\section{HOD Modeling of the Clustering of Faint Red Galaxies}
\label{sec:hod}

With the pairwise volume-limited 2PCF measurements for the $-19<M_r<-18$ 
faint red galaxy 
sample, we perform HOD modeling to study the relationship between faint
red galaxies and
dark matter haloes and discuss the implications for faint red galaxy
formation.
In our modeling, a halo is defined to have a mean density
200 times that of the background universe. We adopt a spatially flat cosmology
with the following cosmological parameters, $\Omega_m=0.25$, $\Omega_b=0.043$,
$n_s=0.95$, $\sigma_8=0.8$, and $h=0.7$.

\subsection{HOD Parameterization}
\label{sec:hodpar}

It is useful and physically meaningful to separate central galaxies and
satellite galaxies in the HOD parameterization \citep[][]{Kravtsov04,Zheng05}.
We follow Z11 to construct the mean occupation function for the faint
red galaxies as 
follows. Since we are modeling galaxies in a luminosity bin, the mean 
occupation function of central galaxies can be obtained by differencing
those for two luminosity-threshold samples. The mean occupation function of
central galaxies for a luminosity-threshold sample can be modeled as
a softened step function \citep{Zheng07},
\begin{equation}
\label{eqn:Ncen}
\langle N_{\rm cen}(M) \rangle =\frac{1}{2}\left[1+{\rm erf}\left(\frac{\log M-\log M_{\rm min}}{\sigma_{\log M}}\right)\right],
\end{equation}
where $M$ is the halo mass, ${\rm erf}$ is the error function, $M_{\rm min}$
is the characteristic minimum mass of haloes that can host the galaxies above
the luminosity threshold, and $\sigma_{\log M}$ characterizes the transition
width of the softened step function (reflecting the scatter in the central 
galaxy luminosity and halo mass). We make use of the HOD modeling results
in Z11 for the $M_r<-18$ and $M_r<-19$ sample and take the difference of the
two corresponding central galaxy occupation functions. The shape (not the
amplitude) of this difference central occupation function is used for the 
central occupation function for the $-19<M_r<-18$ faint red galaxies. 
The exact shape of the mean occupation function of central galaxies 
does not affect the modeling. This is because the role of the central 
galaxies is to contribute to the large scale bias and the number density.
Since the mass of their host haloes is around a few times $10^{11}\hinvMsun$, 
where the halo bias reaches a plateau, the contributions to galaxy bias and
number density from central galaxies are not sensitive to the shape of the 
mean occupation function.

\begin{figure*}
\includegraphics[width=1.0\textwidth]{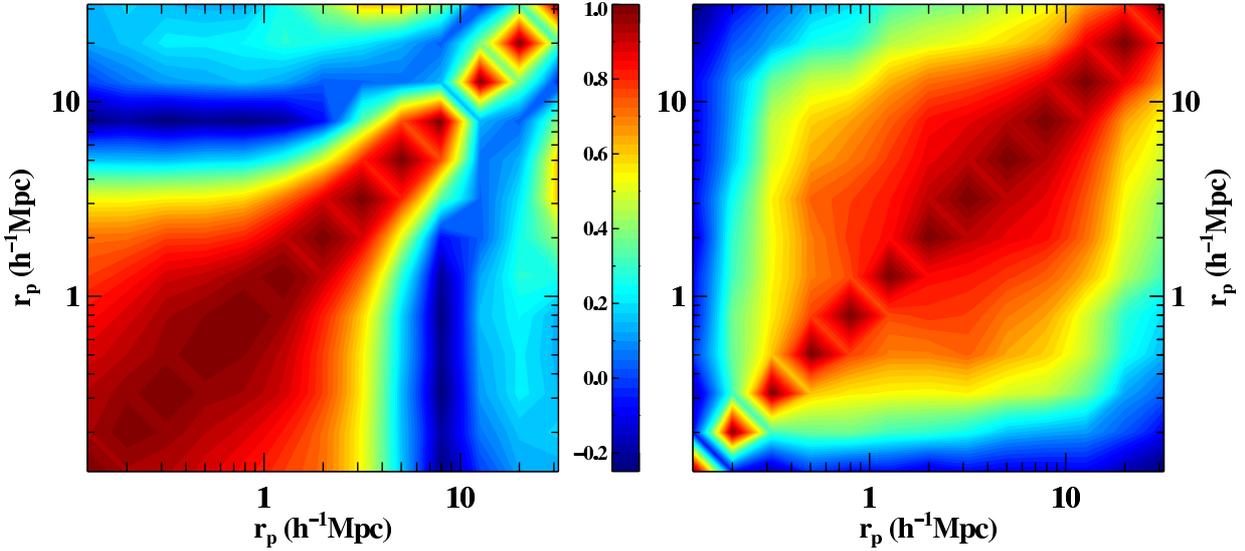}
\caption{
Covariance matrices of $w_p$ for the $-19<M_r<-18$ red galaxies (left) and
the $M_r<-21$ sample (right). While the measurements for
the bright sample are correlated on relatively large scales,
those for the faint sample are highly correlated on small scales, 
suggesting a difference
in the way that galaxies occupy haloes. See text for details.
}
\label{fig:covar}
\end{figure*}

For the satellite galaxies, we model the occupation function as a modified 
power law
\begin{equation}
\label{eqn:Nsat}
\langle N_{\rm sat}(M)\rangle = \left(\frac{M-M_0}{M_1^\prime}\right)^\alpha.
\end{equation}
The satellite occupation number is assumed to follow a Poisson distribution
with the above mean \citep{Kravtsov04,Zheng05}. We find that change in $M_0$ 
does not affect our main results, so we first fix it at the value from the 
$M_r<-18$ model (Z11) and follow Z11 to further modify $N_{\rm sat}(M)$ by 
the $M_r<-18$ central occupation function. So our default model has 
$M_1^\prime$ and $\alpha$ as free parameters, which characterize the 
amplitude and slope of the power law.

We use the shape from the above 
parameterization for the mean occupation functions of central and 
satellite galaxies, and the overall normalization of the mean occupation 
function for each set of HOD parameters is determined by the faint
red galaxy number 
density. We start from assuming that the spatial distribution of satellites 
inside haloes follows the Navarro-Frenk-White (NFW) profile \citep{NFW96},
with the halo concentration parameter $c(M)=c_0(1+z)^{-1}(M/M_{\rm nl})^\beta$
\citep{Bullock01,Zhao09}, where $c_0=11$, $\beta=-0.13$, and 
$M_{\rm nl}=2.26\times10^{12}\hinvMsun$ is the non-linear mass scale 
at $z=0$. Later in this section, we also show results with such  
an assumption relaxed.

The theoretical galaxy 2PCF in our model is calculated using the method 
described in \cite{Zheng04} and \cite{Tinker05}. In addition, we also 
incorporate the effect of residual redshift-space distortions when 
computing $w_p$ from integrating the redshift-space 2PCF up to $\pi_{\rm max}$ 
(instead of infinity), by applying the multipole expansion method of 
\citet{Kaiser87} (also see \citealt{Bosch13}). A Markov Chain Monte Carlo 
(MCMC) method is adopted to explore the HOD parameter space constrained by 
the projected 2PCF $w_p(r_p)$.

\subsection{Modeling Results}

\begin{figure*}
\includegraphics[width=1.0\textwidth]{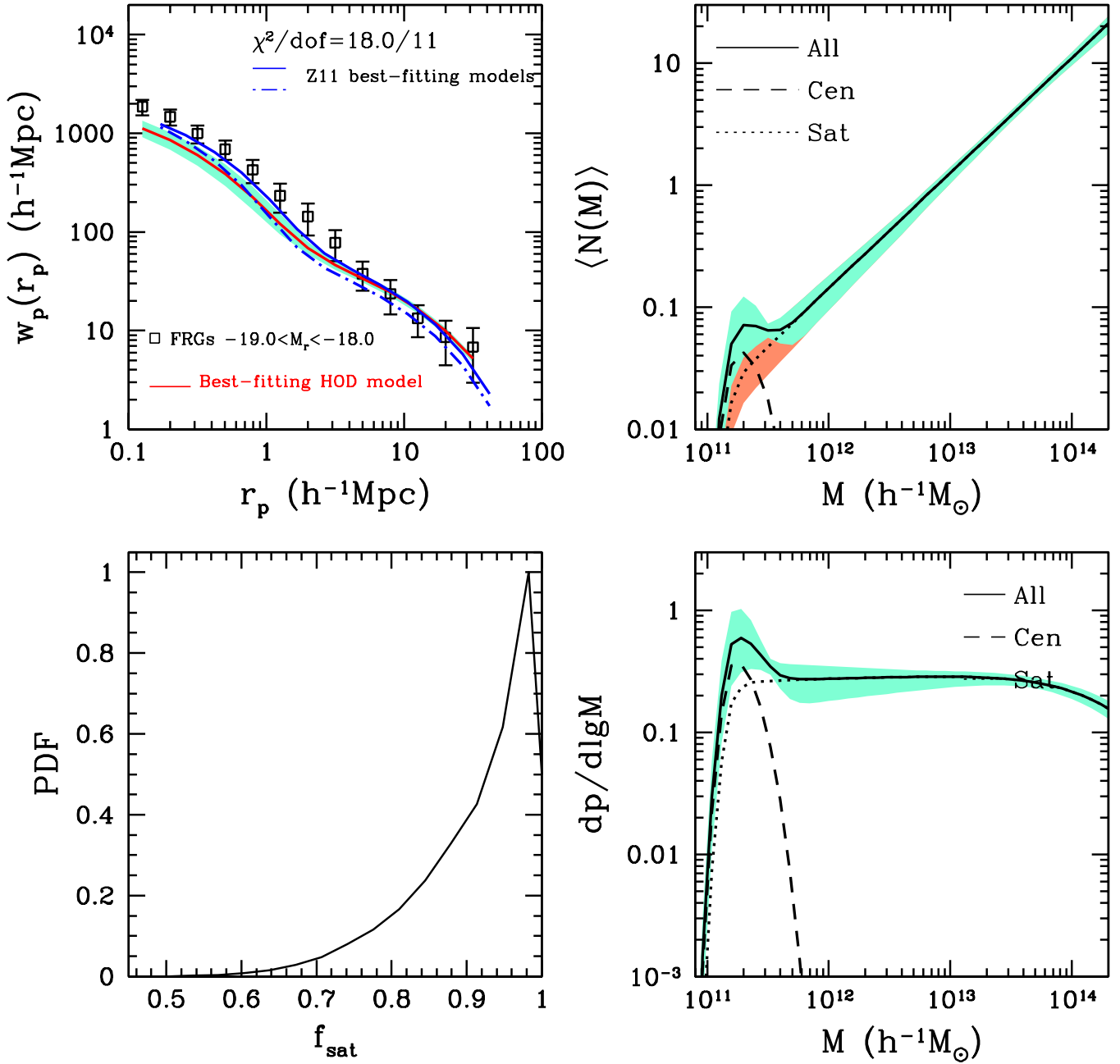}
\caption{
Results from modeling the pairwise volume-limited measurement of the 
projected 2PCF for the $-19<M_r<-18$ red galaxies. Full covariance
matrix is used, and the default HOD model with two parameters are applied.
{\it Top-left:} The measurement and the bestfit, with the shaded region
indicating the range of model curves with the 68.3 per cent lowest values of 
$\chi^2$. For comparison, we overplot two best-fitting HOD models from Z11
as solid and dot-dashed blue curves (see text for details).  
{\it Top-right:} The inferred mean occupation functions, with the shaded
region having the similar meaning as above.
{\it Bottom-right:} The probability distribution of mass of the host haloes,
obtained by multiplying the curve in the top-right panel with the differential
halo mass function.
{\it Bottom-left:} The marginalized distribution of the fraction of faint 
red galaxies
being satellites. 
}
\label{fig:fullcovar}
\end{figure*}

Figure~\ref{fig:fullcovar} shows the modeling result using the full covariance
matrix (also shown as the {\it 2-par HOD model} in Figure~\ref{fig:sam}). The 
default model leads to a reasonably good fit to the data. The 
best-fitting model has a value of $\chi^2=18.0$ for 11 degrees of freedom 
($dof$), which come from 13 $w_p$ data points and 2 free parameters 
($M_1^\prime$ and $\alpha$). Note that the number density $n_g$ is not regarded
as a data point but is only used to normalize the mean occupation function. 

By eye, the best-fitting $w_p$ works well on scales larger than $4\hinvMpc$,
but under-predicts the data on small scales, where the one-halo term dominates.
However, given that the data points on small scales are highly correlated
(Fig.~\ref{fig:covar}), the deviation is not significant. We present the HOD 
constraints in other three panels. The mean occupation function (top-right 
panel) shows a bump for central galaxies around $\log [M/(\hinvMsun)]=11.3$ 
and a power law for satellite galaxies. The curve weighted by the differential 
halo mass function $dn/d\log M$ is shown in the bottom-right panel 
($dp/d\log M = \langle N(M) \rangle dn/d\log M$), and we see that a randomly 
selected satellite faint red galaxy has comparable possibilities to reside in 
any haloes above a few times $10^{11}\hinvMsun$. The central galaxies
has a large contribution to the galaxy bias factor and hence the 2PCF
on large scales. The small-scale 2PCF is mainly contributed by 
satellite-satellite pairs in massive haloes. The fraction of the faint
red galaxies being satellites is above 70 per cent (bottom-left panel), even though 
such a high satellite fraction still makes the model $w_p$ lie below the data 
points on small scales.

In the {\it top-left} panel of Figure~\ref{fig:fullcovar}, we also show as blue
curves the predicted $w_p$ from two best-fitting HOD models in Z11 (see the
top-left panel of their Figure 21). The 
solid blue curve corresponds to the case with an HOD parameterization similar 
to ours. We find that their best-fitting model is able to reproduce our 
improved $w_p$ measurements on large scales and the corresponding mean 
occupation function and satellite fraction ($\sim$90 per cent) is consistent 
with our constraints. It is interesting that this Z11 best-fitting HOD model 
``predicts'' our
improved $w_p$ measurements on large scales, even though it is based on the 
original measurements with much lower amplitudes. The reason for this is that
the small-scale $w_p$ plays a significant role in determining the HOD. The 
dot-dashed blue curve is from an alternative model in Z11, where satellites
are only allowed to populate massive haloes (above $10^{13}\hinvMsun$) and 
there are more central galaxies in low-mass haloes in order
to better match the low-amplitude large-scale $w_p$ measurements in Z11. The
corresponding mean occupation function of central galaxies (not shown in 
Figure~\ref{fig:fullcovar} to avoid crowding) is about a factor of 5 higher 
than our best-fit. Clearly this alternative model underpredicts our improved 
large-scale $w_p$ measurements.


To investigate the robustness of the HOD constraints, especially the 
high satellite fraction, we perform two model tests by exploring
two possibilities, inaccurate covariance matrix and insufficient
model. 

Given the low luminosity nature of the sample under study, the survey volume is
small, $\sim$$1.5\times 10^6 h^{-3}{\rm Mpc}^3$ for the faint-end 
volume-limited sample (e.g. $\sim$30 times smaller than that of the
$L>L_*$ sample). Even though our pairwise volume-limited measurements
effectively increase the volume by a factor about two, it is still small
(i.e., roughly corresponding to a cubic box with 150$\hinvMpc$ on a side). 
The covariance matrix estimated from the small volume may not be accurate.
Noises in the covariance matrix can affect the value of $\chi^2$ and the
constraints. Compared
to the covariance between data points at different values of $r_p$, the
diagonal elements of the covariance matrix are more reliably determined. We 
perform a test by using only the diagonal elements in the fit. 

A perfect fit is achieved, with the best-fitting $\chi^2/dof=4.5/11$. 
The fit to small-scale data points is improved by populating more satellites
into more massive haloes, i.e., a higher high-mass end slope $\alpha$ of the
satellite mean occupation function. A larger fraction of 
satellites reside in haloes above $10^{13}\hinvMsun$.
Compared to the fit with the full covariance matrix, the overall satellite 
fraction shows a bimodal distribution, peaked around $\sim$25 per cent and 
$\sim$70 per cent. This test with
diagonal errors of the covariance matrix shows that the noises in the 
covariance matrix may affect the best-fitting $\chi^2$, but satellites
in massive haloes are necessary to explain the strong small scale clustering.
Since this model neglects the strong correlation among small-scale data
points, the results should be taken with a grain of salt. 

\begin{figure}
\includegraphics[width=0.45\textwidth]{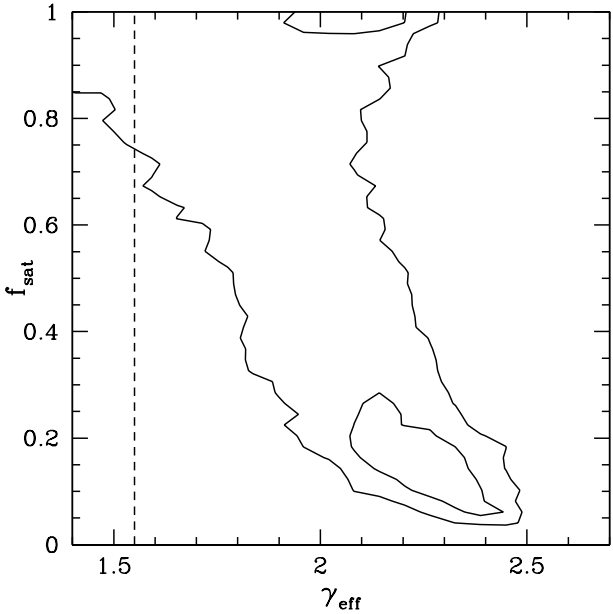}
\caption{
Marginalized distribution of satellite fraction and effective slope of
inner profile of satellite distribution. The effective slope of a generalized
NFW profile is computed at $r=0.1\hinvMpc$ in haloes of mass 
$2 \times 10^{14}\hinvMsun$. See text for detail.
The contours represent the 68.3 per cent and 95.5 per cent confidence levels.
The vertical dash line marks $\gamma_{\rm eff} = 1.55$, the value corresponding
to slope of the NFW profile at this radius. 
}
\label{fig:fsat_gamma}
\end{figure}

Next, we add more flexibility in the model by allowing the spatial distribution
of satellites to deviate from the NFW profile. A generalized NFW profile is
adopted, with the concentration parameter $c$ and inner slope $\gamma$ as 
two additional free parameters \citep[e.g.][]{Watson10,Watson12,Bosch13,Guo14},
\begin{equation}
\label{eqn:gNFW}
\rho(r)\propto \left[ \left( \frac{cr}{r_{\rm vir}}\right)^\gamma 
\left(1+\frac{cr}{r_{\rm vir}}\right)^{3-\gamma}\right]^{-1},
\end{equation}
where $r_{\rm vir}$ is the halo virial radius.
The modeling 
results with the full covariance matrix is shown in 
Figure~\ref{fig:sam}, termed as the {\it 4-par HOD model}. The best-fitting 
$\chi^2=14.9/9$ is 
reasonable. The model still under-predicts the small-scale clustering, but these
data points are highly correlated. The mean occupation function is similar 
to the case with diagonal covariance matrix, but with a much large spread. 
The fit shows a bimodal distribution, favoring either a high satellite
fraction ($>$70 per cent) or a steeper inner profile. To illustrate this, in
Figure~\ref{fig:fsat_gamma}, we show the marginalized distribution of 
the derived satellite fraction $f_{\rm sat}$ and the effective inner slope 
$\gamma_{\rm eff}$. We choose to compute the effective slope at 
$r=0.1\hinvMpc$ in haloes of mass $2 \times 10^{14}\hinvMsun$,
\begin{equation}
\gamma_{\rm eff}\equiv -\frac{d\ln \rho(r)}{d\ln r} 
= \gamma +\frac{c(3-\gamma)}{c+r_{\rm vir}/r}.
\end{equation}
For comparison, the vertical dashed line shows the effective slope for the
corresponding NFW profile ($\gamma=1$).

The steeper 
inner profile helps enhance the small-scale clustering, and thus reduces 
the need for a high satellite fraction (only need $f_{\rm sat}\sim$20 per cent).
If the diagonal covariance matrix were used with the generalized NFW profile, 
the fit would have best-fitting $\chi^2/dof=2.0/9$, and
the satellite fraction is close to a uniform distribution from $\sim$20 per cent to 
$\sim$100 per cent.

If we further allow the cutoff mass $M_0$ in the satellite mean occupation 
function to float, the
way to populate satellites becomes even more degenerate, ranging from 
solutions similar to the above cases to those with satellite galaxies 
only populating very massive haloes (e.g. above $10^{13.5}\hinvMsun$). 

From the above modeling results and various tests, we find that the satellite 
fraction is degenerate with the spatial distribution profile of satellites.
For a profile not far from NFW, the strong small-scale clustering requires 
a large fraction of the faint red galaxies to be satellite galaxies in high 
mass haloes. Even with steep inner profiles, a fraction of faint red galaxies
are still needed to be satellites in high mass haloes.

Breaking the degeneracy between the high satellite fraction and the steep
inner profile needs additional information. The pattern seen in the
covariance matrix (left panel of Figure~\ref{fig:covar}) in fact supports 
the solution that at least a considerable fraction of faint red galaxies need
to be satellites in massive haloes.
The strong 
correlation occurs for data points that are dominated by contributions from
intra-halo satellite-satellite pairs in massive haloes, so the clustering
amplitude on such scales is sensitive to the number of massive haloes within
the given (survey) volume. The covariance matrix reflects the covariance of 
the measurements across different regions (of the same volume) in the universe 
(or `same' region in different realizations of the universe). Such a sample 
variance effect affects more the number of massive haloes than that of low 
mass haloes. That is, the number of massive haloes fluctuates more than that of
low-mass haloes across different regions. In slightly denser regions, we 
expect to have more haloes of higher masses, which increases the number of
intra-halo satellite-satellite pairs and thus enhances the small-scale 
clustering (over all relevant one-halo scales simultaneously). In regions 
of lower density, we have the opposite effect that the small-scale clustering
becomes weaker simultaneously. This explains the relatively strong correlation
among small-scale data points. On the contrary, the large-scale clustering
has a large contribution from faint red galaxies residing in haloes of a few 
times $10^{11}\hinvMsun$ (as central galaxies). The abundance of such 
low-mass haloes does not fluctuate
much across different regions of the above volume, and therefore
a strong correlation is not seen among the large-scale data points.

\begin{figure}
\includegraphics[width=0.5\textwidth]{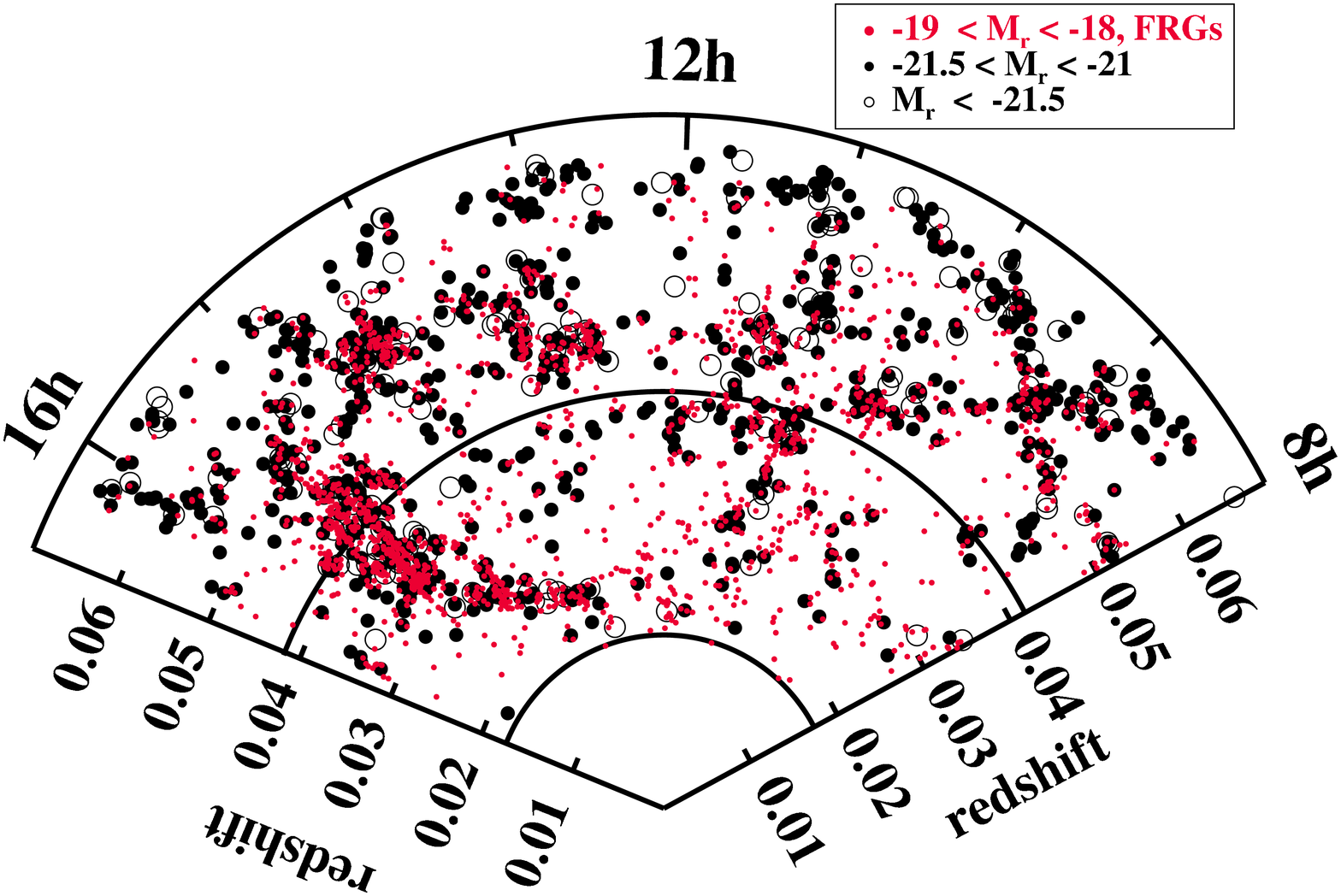}
\caption{
Comparison between the spatial distributions of faint red galaxies and
more luminous galaxies. The slice is chosen to be around the celestial 
equator with declination in the range of $|\delta|< 10^{\circ}$.
The two inner arcs ($0.017<z<0.042$) marks the boundary for the faint-end 
volume-limited $-19<M_r<-18$ sample.
}
\label{fig:pie}
\end{figure}

\begin{figure}
\includegraphics[width=0.45\textwidth]{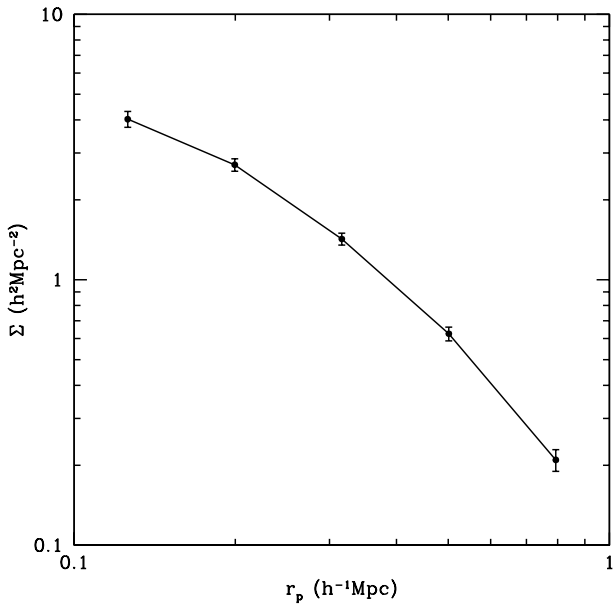}
\caption{ 
Mean excess surface density of faint red galaxies ($-19<M_r<-18$) around 
one luminous galaxies with $M_r<-21$. Poisson error bars are plotted.
}
\label{fig:fbcross}
\end{figure}

A considerable fraction of the faint red galaxies being satellites 
in massive haloes can help
explain the high pairwise velocity dispersion inferred from small scales 
\citep{Li06b} and small-scale redshift-space distortions \citep{Ross11}.
The need for satellites is also supported by a visual check of the
spatial distribution in a slice of volume, as shown in Figure~\ref{fig:pie}. 
The red dots are $-19<M_r<-18$ red galaxies. The filled and open black circles
are more luminous galaxies, displaying the locations of high mass haloes
($\log M \ga 12.8$ for $M_r<-21$ and $\log M\ga 13.4$ for $M_r<-21.5$; see Z11).
A large fraction of the faint red galaxies are found close to these high mass 
haloes residing in over-dense regions, suggesting a high probability of being 
satellites. However, there are also faint red galaxies in the `void' regions 
in the diagram, which means that a fraction of them can be central galaxies of 
low mass haloes. This is consistent with the mean occupation function having
both central and satellite components, as in our HOD model.

The visual impression of a high probability of faint red galaxies being 
satellites in Figure~\ref{fig:pie} can be further justified by counting
the correlated pairs of faint and luminous galaxies, which also helps
remove the confusion from chance alignment. Figure~\ref{fig:fbcross} shows
the mean excess surface density of faint red galaxies ($-19<M_r<-18$) 
per luminous galaxy with $M_r<-21$ as a function of the projected distance
to the luminous galaxy, from counting faint red galaxies within 
$\Delta \pi=\pm 30\hinvMpc$ of luminous galaxies. This is done in the volume 
where the faint galaxies are complete and is equivalent to the 
cross-correlation function between faint red and luminous galaxies. Since the 
majority of $M_r<-21$ galaxies are central galaxies ($\sim 91$ per cent; 
see Z11), the excess faint red galaxies are almost all satellite galaxies.
From the above density profile and the number density of luminous galaxies, 
we can infer the number density of faint red galaxies within a projected
distance 1$h^{-1}{\rm Mpc}$ away from luminous galaxies, which is about 56
per cent of all faint red galaxies. As a rough estimate of the fraction
of (satellite) faint red galaxies in host haloes of $M_r<-21$ galaxies,
this number is close to the satellite fraction (about 45 per cent) in 
$M_r<-21$ host haloes inferred from our best-fitting HOD model of faint 
red galaxies and the halo mass distribution of $M_r<-21$ galaxies (Z11).
Our result that most faint red galaxies are satellites is therefore well
supported.

 
\begin{figure*}
\includegraphics[width=1.0\textwidth]{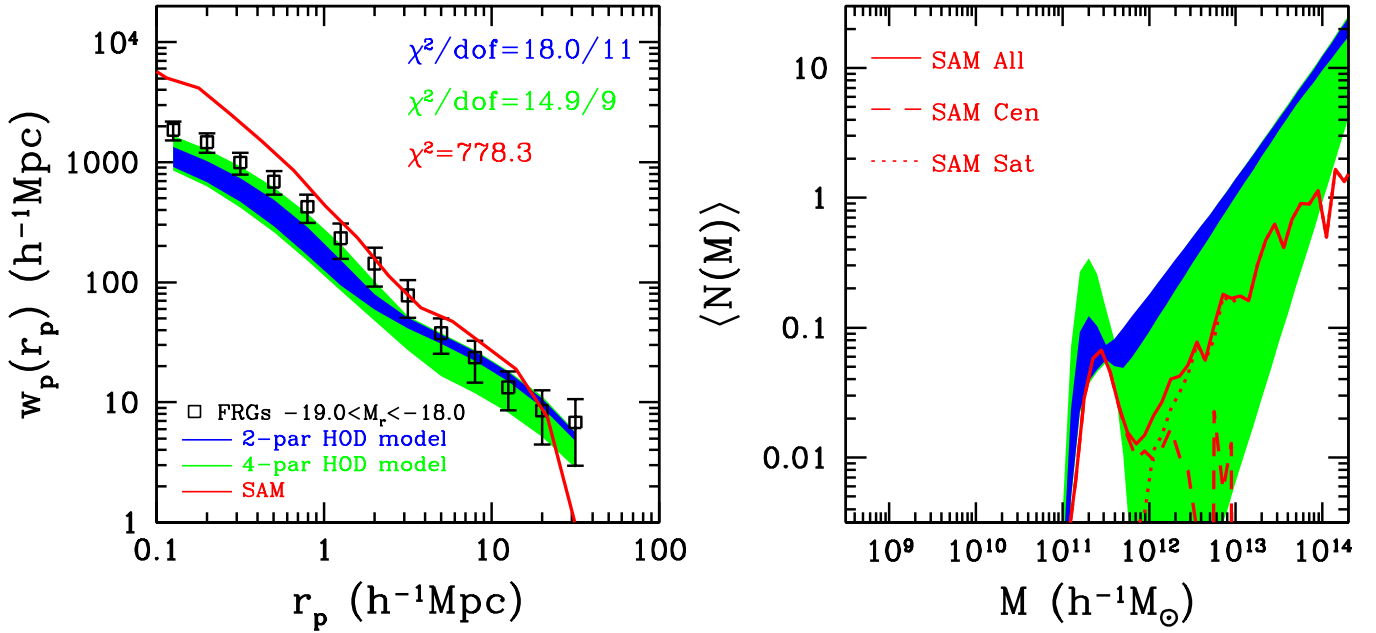}
\caption{
Comparison between two HOD modeling results and the SAM prediction for
the clustering and mean occupation functions of the $-19<M_r<-18$ red 
galaxies. The {\it 2-par HOD model} has the amplitude ($M_1^\prime$) 
and slope ($\alpha$) in the power law of satellite mean occupation function
as free parameters, while the {\it 4-par HOD model} adopts a generalized
NFW profile for satellite distribution inside haloes, adding the 
concentration ($c$) and inner density slope ($\gamma$) as two additional
parameters. See text in detail. Green and blue shaded region shows the
envelopes from models with the 68.3 per cent lowest values of $\chi^2$.  
}
\label{fig:sam}
\end{figure*}

Finally, we compare our modeling results with predictions from the 
semi-analytic model (SAM) of galaxy formation in \citet{Guo13}. More 
specifically, we construct the faint red galaxies sample with the MS-SW7 SAM catalog 
(WMAP7 cosmology), by adopting the same luminosity and colour cut as used 
in defining our faint red galaxy sample. The mock sample has a number density of 
$1.46\times 10^{-3} h^3{\rm Mpc}^{-3}$, about 40 per cent that of the observed 
sample. The SAM predicted $w_p$ is shown in the left panel of 
Figure~\ref{fig:sam}, together with the range of model fits from the
default 2-parameter and the 4-parameter HOD models. On scales above
1$h^{-1}{\rm Mpc}$, the prediction falls on top of the data. However,
the SAM clearly over-predicts $w_p$ below 1$h^{-1}{\rm Mpc}$, as noted 
by \citet{Guo13}. The SAM prediction also appears to have a steeper slope 
than the data points. To be quantitative, 
the $\chi^2$ calculated from comparing the SAM curve with the 
data points is $778.3$, which is a poor fit.

\begin{figure}
\includegraphics[width=0.45\textwidth]{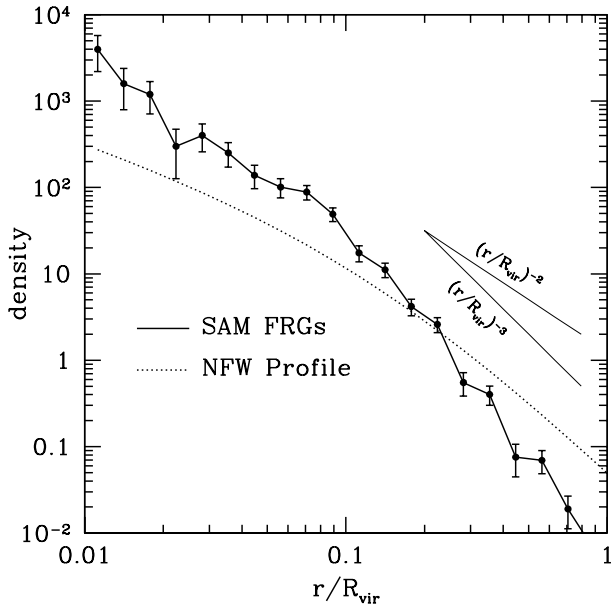}
\caption{
Comparison between the spatial distribution of SAM faint red satellites and
the NFW profile in massive haloes. Both profiles are obtained by stacking 
haloes above $10^{13}\hinvMsun$ with radius normalized to halo virial radius.
Two straight solid lines mark two power-law profiles with different
slopes to guide the eye.
}
\label{fig:profcmp}
\end{figure}

The SAM predicted mean occupation function is shown in the right panel of
Figure~\ref{fig:sam}. Encouragingly, it has the similar central component
(a bump around $\log M=11.3$) and satellite component (roughly following 
a power law) as inferred from the HOD modeling. The satellite mean occupation
function has a similar slope as in our default model with two parameters 
(blue). It then implies that the SAM would predict a similar small-scale slope 
in $w_p$ as the HOD model. However, the SAM leads to a steeper slope as seen
in the left panel, which means that the inner spatial distribution of SAM 
satellites is too steep. In Figure~\ref{fig:profcmp}, we compare the spatial 
distribution of SAM satellites and the NFW profile, stacked with haloes of 
masses above $10^{13}\hinvMsun$ and radius normalized to halo virial radius. 
The scales relevant to the measurements ($r_p>0.1\hinvMpc$) correspond to 
$r/R_{\rm vir}\ga 0.2$, and clearly the SAM satellites show a much steeper
profile than the NFW profile used in our modeling.
We also find that the SAM predicts a satellite fraction 
of 35 per cent, and it lies in the middle of the two extreme solutions with the
4-parameter HOD model (in between high satellite fraction and very steep
inner profile). The steep rise in $w_p$ towards small scales from SAM
suggests that its predicted combination of satellite fraction and inner 
distribution profile deviates from that implied by the data. 

As for the physical origin of the over-prediction of the small-scale 
clustering by the SAM model, \citet{Guo13} attribute it to the over-abundance
of low-mass galaxies and/or galaxies forming too early. 
This is supported by the results that the stellar mass and luminosity 
functions of satellites and the radial distribution of satellites 
around primary galaxies in the SAM model match those observed in the SDSS,
while large discrepancies show up when split by color, with red
satellites being over-predicted \citep{Wang12,Wang14}, pointing the origin
to the quenching of satellites. Modifying baryon processes and 
environmental effects can make significant improvements \citep{Henriques15}.
For the particular faint red galaxy population we study here, the SAM 
model underpredicts the satellite fraction, compared to the HOD modeling 
results.

From such a comparison,
the steeper slope in the spatial distribution 
of SAM satellites indicates that the SAM implementation may shut off star
formation in faint satellites too late in massive haloes, making them more
concentrated toward the halo centre, or that the efficiency of disrupting
faint red galaxies toward the halo centre or merging them with the central 
galaxy is too low in the SAM model.

\section{Summary and Discussion}
\label{sec:discussion}

We introduce a method to improve the volume-limited 2PCF measurements of 
faint galaxies by making a full use of the flux-limited sample. The method is
then applied to the $-19<M_r<-18$ SDSS red galaxies to obtain a more accurate
projected 2PCF measurement, and we perform HOD modeling to infer the relation
between these faint red galaxies and dark matter haloes.

The depth of a galaxy redshift survey limits the observation of faint 
galaxies, which cannot be observed to large redshifts. For the ease of
interpretation, the clustering of faint galaxies is usually studied with 
a small, volume-limited sample, giving rise to considerable sample 
variance. The method introduced in this work (termed as `pairwise 
volume-limited measurement'), when applied to luminosity-bin
or luminosity-threshold galaxy samples, decomposes the galaxy pair 
contribution to
correlations of pairs with same or different luminosities. Each pair count
is made volume-limited, by weighing with the appropriate volume factor
(i.e. common volume of being complete for both galaxies). It shares some
similarity with the $1/V_{\rm max}$ method to estimate galaxy luminosity 
function. The final form of the method is a generalized Landy-Szalay 
estimator. The method uses all possible pairs in a flux-limited sample to form 
an effectively volume-limited measurements. With the contribution to the 2PCF 
from more luminous galaxies in the sample measured in a larger volume, we 
effectively increase the sample volume (by a factor of about two). As a 
consequence, the uncertainty in the 2PCF measurements can be improved (by tens 
of per cent) and the effect of sample variance is reduced. Since galaxy 
luminosity function drops exponentially above $L_*$, the method turns out to be
most efficient for samples of faint galaxies (with luminosity below $L_*$).

We apply the method to the $-19<M_r<-18$ SDSS red galaxies and make the
pairwise volume-limited projected 2PCF measurements. Compared to those
from the commonly used volume-limited sample, the improved measurements have
$\sim$40 per cent smaller uncertainties on large scales. Furthermore, the 
previous puzzling result that the sample shows a strong small-scale clustering
and a weak large-scale clustering (Z11) is not seen in the improved 
measurements. Suffering less from the effect of sample variance, the measured 
$w_p$ shows a more consistent trend from small to large scales. 
In addition, we also apply the method to all the $-19<M_r<-18$ galaxies 
(red+blue) and measure the large-scale galaxy bias factor from the ratio
of the measured $w_p$ to that of the dark matter. Compared to 
the bias factor $0.90\pm0.10$ inferred in Z11 (left panel of their Fig.7), 
we find the updated value to be $0.99\pm0.12$, much more consistent with the 
value derived from HOD modeling.

With the previous volume-limited measurements, it was difficult for HOD modeling to simultaneously reproduce the strong small-scale clustering
and the weak large-scale clustering. As a possible explanation of the data,
Z11 invoke a scenario to put satellites only in haloes above 
$10^{13}\hinvMsun$ and have central galaxies in haloes of a few times
$10^{11}\hinvMsun$. The motivation is that gas accretion and star 
formation in satellites shut off more efficiently in haloes more massive than
the previous host haloes of those satellites before they are accreted into 
the haloes \citep[e.g.][]{Font08,Kang08,Simha09}. With our improved 
measurements, in particular those on large scales, we find that the above
scenario is not necessary in explaining the data with our simple two-parameter
model, but remains a possibility for the more flexible model. With the 
two-parameter model, a reasonable fit to the data can be achieved and we 
find that the faint red satellites can reside in haloes ranging 
from a few times $10^{11}\hinvMsun$ to cluster-size haloes more massive 
than $10^{14}\hinvMsun$, leading to large satellite fraction ($>$70 per cent).
Unsurprisingly, with the more flexible, four-parameter model, we find that a 
steep inner profile of the spatial distribution of satellites inside haloes 
can compensate the high satellite fraction. However, either solution needs 
to put a fraction of satellites in massive haloes. Such a picture is supported 
by the strong covariance among small-scale data points and by the relative 
spatial distribution of the faint red galaxies and more luminous galaxies 
tracing massive haloes.
 
The mean occupation function of those faint red galaxies predicted in the 
SAM model of \citet{Guo13} falls into the range of our model constraints.
However, the predicted small-scale clustering appears to be too strong to be 
consistent with the measurements, indicating a too steep inner distribution
profile of satellites in the SAM model. It suggests that the disruption
or merging of faint red galaxies in massive haloes are more 
efficient than that implemented in the SAM model.

Although we develop a method to make more accurate volume-limited measurements
of faint galaxy clustering, we still lack very tight HOD constraints and
could not draw too strong a conclusion regarding the nature of the faint
red galaxies. This limits the power of the 2PCF measurements to efficiently 
assess the effects like assembly bias \citep[e.g.][]{Zentner14}. 
The role of assembly bias in the clustering of faint red galaxies deserves
a further investigation.
Combining the 2PCFs with other statistics, like those from group catalogs 
\citep[e.g.][]{Skibba09,Wang09}, can be helpful. A deep and large survey 
probably is the ultimate resort to using clustering to study the faint galaxy 
population.

\section*{Acknowledgements}

We thank Cheng Li and David Weinberg for helpful discussions and Simon White
for useful comments. 
HJX thanks Yiping Shu for technical help. ZZ is partially supported 
by NSF grant AST-1208891 and NASA grant NNX14AC89G. IZ and ZZ also 
acknowledge NSF grant AST-0907947.

Funding for the Sloan Digital Sky Survey (SDSS) has been provided by the Alfred P. Sloan Foundation, the Participating Institutions, the National Aeronautics and Space Administration, the National Science Foundation, the U.S. Department of Energy, the Japanese Monbukagakusho, and the Max Planck Society. The SDSS Web site is http://www.sdss.org/.

The SDSS is managed by the Astrophysical Research Consortium (ARC) for the Participating Institutions. The Participating Institutions are The University of Chicago, Fermilab, the Institute for Advanced Study, the Japan Participation Group, The Johns Hopkins University, Los Alamos National Laboratory, the Max-Planck-Institute for Astronomy (MPIA), the Max-Planck-Institute for Astrophysics (MPA), New Mexico State University, University of Pittsburgh, Princeton University, the United States Naval Observatory, and the University of Washington.

\appendix 
\section{Generalized Landy-Szalay Estimator for Pairwise Volume-limited Galaxy 2PCF Measurement}
\label{sec:appendix1}

Here we derive the expression for the pairwise volume-limited galaxy 2PCF 
measurement from a flux-limited galaxy sample.

First, consider a volume-limited sample of galaxies in a luminosity bin 
[$L_0$, $L_N$]. We divide it into $N$ subsamples of fine luminosity bins, 
[$L_0$, $L_1$], [$L_1$, $L_2$], ..., [$L_{N-1}$, $L_N$] ($L_0<L_1<...<L_N$). 
The 2-point auto-correlation function $\xi$ of the whole 
sample can then be decomposed into contributions from the the 2-point 
auto-correlation function $\xi_{ii}$
($i=1,2,...,N$) of each subsample and the 2-point cross-correlation 
functions $\xi_{ij}$ ($i,j=1,2,...,N$; $i\neq j$) among different subsamples
\citep{Zu08}, 
\begin{equation}
\label{eqn:xi_decomp}
\xi=\sum\limits_{i,j}\frac{\bar{n}_i\bar{n}_j}{\bar{n}^{2}}\xi_{ij},
\end{equation}
where $\bar{n}_i$ is the mean number density of the $i$-th subsample and
$\bar{n}=\sum \bar{n}_i$ is that of the whole sample.
This is just a reflection that the total number of pairs is the sum over
all possible auto-pairs ($i$-$i$) and cross-pairs ($i$-$j$). Note that 
in the summation, both $i$ and $j$ go from 1 to $N$, hence no coefficient 
2 in front of the cross-correlation terms.

For a flux-limited sample of galaxies with the same luminosity range 
[$L_0$, $L_N$], we can do the same division into subsamples of fine 
luminosity bins.
Each subsample has its own corresponding maximum volume that makes it
volume-limited, i.e., the subsample of galaxies are complete within this 
volume. Denote $V_i$ as such a volume for the $i$-th subsample. To compute
the 2PCF of galaxies in the [$L_0$, $L_N$] bin in a volume-limited sense, 
equation~(\ref{eqn:xi_decomp}) 
still applies, but we can evaluate $\xi_{ij}$ in a volume that both the 
$i$-th subsample and $j$-th subsample are complete. Denote the common volume
as $V_{ij}$. Obviously $V_{ij}={\rm min} \{ V_i, V_j \}$, determined
by the fainter subsample of the two. The correlation function $\xi_{ij}$ can be
calculated using the Landy-Szalay estimator in the volume $V_{ij}$,
\begin{equation}
\label{eqn:LSij}
\xi_{ij}=\left. \frac{{\rm DD}_{ij}-2{\rm DR}_{ij}+{\rm RR}_{ij}}
                     {{\rm RR}_{ij}}\right\vert_{V_{ij}},
\end{equation}
where ${\rm DD}_{ij}$, ${\rm DR}_{ij}$, and ${\rm RR}_{ij}$ are the
numbers of $i$-$j$ data-data, data-random, and random-random pairs. 
For measuring the correlation function at pair separation $\vecr$ 
[e.g. $(r_p,\, \pi)$], each of the 
above pair counts should be understood as the number of pairs with separation 
within a bin $d\vecr$ around $\vecr$ (i.e., $\vecr\pm d\vecr/2$).
For the simplicity of the derivation later, we assume the same number of data 
and random points, so we do not need to normalize those pair counts. 
We also note that the random catalog should be constructed in a way that each 
random point is assigned a luminosity, drawn randomly from the luminosity 
distribution of the galaxy sample. That is, at any small luminosity
bin, random points have the same redshift distribution and volume of 
completeness as the galaxy sample.

Now let us take a look at the weight in equation~(\ref{eqn:xi_decomp}), which
is determined by the subsample number densities. We can link the number 
densities to pair counts from the random catalog. For pairs composed of 
random points in the $i$-th and $j$-th luminosity bins, we can compute the
counts $\left. {\rm RR}_{ij}(\vecr)\right\vert_{V_{ij}}$ of $i$-$j$ pairs with 
separation $\vecr\pm d\vecr/2$ in the common volume $V_{ij}$ of completeness. 
Under the condition $r^3\ll V_{ij}$, the counts are simply the product of 
the number of $j$-luminosity points in a volume $dV=d^3\vecr$ with separation
$\vecr$ from each 
$i$-luminosity points and the total number of $i$-luminosity points in volume 
$V_{ij}$, with the former $\bar{n}_j dV$ and the latter 
$\bar{n}_iV_{ij}$, i.e.
\begin{equation}
\left. {\rm RR}_{ij}(\vecr)\right\vert_{V_{ij}} = \bar{n}_i\bar{n}_j V_{ij} dV.
\end{equation}
We then have
\begin{equation}
\label{eqn:ninj}
\bar{n}_i\bar{n}_j = \frac{1}{dV}\frac{1}{V_{ij}} \left. {\rm RR}_{ij}(\vecr)\right\vert_{V_{ij}}.
\end{equation}
From $\bar{n}=\sum \bar{n}_\alpha$, $\bar{n}^2$ can be expressed as
$\bar{n}^2=\sum_{\alpha,\beta} \bar{n}_\alpha\bar{n}_\beta$ 
($\alpha,\beta=1,2,...,N$). Substituting equation~(\ref{eqn:ninj}) into the
expression, we have
\begin{equation}
\label{eqn:nsq}
\bar{n}^2=\frac{1}{dV} \sum\limits_{\alpha,\beta} \frac{1}{V_{\alpha\beta}} \left. {\rm RR}_{\alpha\beta}(\vecr)\right\vert_{V_{\alpha\beta}}.
\end{equation}
Substituting equations~(\ref{eqn:LSij}), (\ref{eqn:ninj}), and (\ref{eqn:nsq}) 
into equation~(\ref{eqn:xi_decomp}) results in
\begin{equation}
\label{eqn:extLS}
\xi(\vecr)= 
\frac{\sum\limits_{i,j}\frac{1}{V_{ij}} \left.\left[{\rm DD}_{ij}(\vecr) - 2{\rm DR}_{ij}(\vecr) + {\rm RR}_{ij}(\vecr)\right]\right\vert_{V_{ij}}
}
{\sum\limits_{i,j}\frac{1}{V_{ij}} \left. {\rm RR}_{ij}(\vecr)\right\vert_{V_{ij}}}.
\end{equation}
This equation provides a generalization to the Landy \& Szalay estimator, which 
can be applied to flux-limited sample to make effectively volume-limited galaxy 
2PCF measurements. The derivation assumes the same number of data and random
points. In practice, the random sample is much larger than the data sample.
In such a case, it is easy to show that one needs to use the normalized pair 
counts, ${\rm DD}_{ij}$, ${\rm DR}_{ij}$, and ${\rm RR}_{ij}$. 

The estimator in equation~(\ref{eqn:extLS}) states that the pair counts are 
weighted by $1/V_{ij}$, where $V_{ij}$ is the common volume that both 
galaxies in the $i$-th and $j$-th luminosity bins are complete. This is 
similar to the $1/V_{\rm max}$ method of estimating the galaxy luminosity 
function from a flux-limited sample. In practice, we do not need to divide 
the sample into finite luminosity bins --- for each pair (data-data, 
data-random, or random-random), if both members of the
pair fall into the common volume $V$ of completeness determined by the
fainter member, we keep the pair and give it a weight $1/V$.
We note that a similar $1/V$ weight to galaxy pairs has been adopted by 
\citet{Li06a}, \citet{Li09} and \citet{Li12} in estimating the distribution
of stellar mass in the universe and the stellar-mass-dependent galaxy 2PCFs.
While $V$ is also defined as the lower $V_{\rm max}$ of the two galaxies of 
each pair, there is a subtle difference between our method and theirs in how 
the weight is applied. They apply the weight to all 
galaxy pairs, including also galaxy pairs with one galaxy inside $V$ and 
the other outside $V$. However, in our method based on a formal derivation of 
the modified LS estimator from the perspective of decomposing the 2PCF, we do 
not count such galaxy pairs in measuring the 2PCFs. 
Our method implicitly assumes that the two galaxies of each correlated pair
are at their respective cosmological distances from their redshifts. This 
becomes less accurate at smaller separations (e.g., with $r_p$ less than a few 
Mpc), where the redshift difference of the pair of galaxies has a large 
contribution from the relative peculiar velocities and we tend to miss some 
correlated pairs. On the other hand, while their method would not miss such 
pairs, it over-counts the number of pairs with large projected separations. 
In practice, 
given the large value of $\pi_{\rm max}$, the fraction of either under-counted
or over-counted pairs is tiny and such a subtle difference has no noticeable 
effect on the measurements.

The estimator we introduced makes a maximal use of 
the galaxy sample to achieve effectively volume-limited 2PCF measurements
that have higher signal-to-noise ratio than those from the corresponding 
volume-limited sample. As discussed in the text, the method is mostly
effective for faint galaxy samples in the power-law part of the luminosity
function, as the number density of galaxies drops slowly towards high
luminosity. For luminous galaxies, the rapid (exponential) decrease of
the luminosity function towards high luminosity end makes the gain in volume
from more luminous galaxies less important, as their contribution to the
2PCF is already small in the first place.

\end{document}